\documentclass[aps,prl,twocolumn,superscriptaddress,showpacs]{revtex4}

\usepackage{graphicx}
\newcommand{\ZCT}{Zn$_{1-x}$Cr$_{x}$Te}
\newcommand{\ZCTfive}{Zn$_{0.95}$Cr$_{0.05}$Te}
\newcommand{\ZCTseven}{Zn$_{0.93}$Cr$_{0.07}$Te}

\newcommand{\N}{$[{\rm N}]$}
\newcommand{\I}{$[\,{\rm I}\,]$}
\newcommand{\Tc}{$T_{\rm C}$}
\newcommand{\thetaP}{$\Theta_{P}$}
\newcommand{\TCdItwo}{$T_{\rm CdI_2}$}
\newcommand{\EF}{$E_{\rm F}$}
\begin{document}


\title{
Significant enhancement of ferromagnetism in Zn$_{\mbox{\bf\small 1-\boldmath{$x$}}}$Cr$_{\mbox{\small \boldmath{$x$}}}$Te doped with iodine as an $\mbox{\boldmath{$n$}}$-type dopant
}

\author{Nobuhiko~Ozaki}
\altaffiliation[Present address: ]{Center for Tsukuba Advanced Research Alliance (TARA), University of Tsukuba, 1-1-1 Tennoudai, Tsukuba, Ibaraki, 305-8577, JAPAN}
\author{Nozomi~Nishizawa}
\author{St\'ephane~Marcet}
\altaffiliation[On leave from ]{Laboratoire de Spectrom\'etrie Physique, Universit\'e Joseph Fourier, F-38402 St. Martin d'H\`eres, FRANCE}
\author{Shinji~Kuroda}
\email[electronic address: ]{kuroda@ims.tsukuba.ac.jp}
\affiliation{Institute of Materials Science, University of Tsukuba, 1-1-1 Tennoudai, Tsukuba, Ibaraki 305-8573, JAPAN }
\author{Osamu~Eryu}
\affiliation{Department of Electrical and Computer Engineering, Nagoya Institute of Technology, Gokiso-cho, Showa-ku, Nagoya 466-8555, JAPAN }
\author{K\^oki~Takita}
\affiliation{Institute of Materials Science, University of Tsukuba, 1-1-1 Tennoudai, Tsukuba, Ibaraki 305-8573, JAPAN }


\date{11 July 2006}

\begin{abstract}
The effect of additional doping of charge impurities was investigated in a ferromagnetic semiconductor {\ZCT}.
It was found that the doping of iodine, which is expected to act as an $n$-type dopant in ZnTe, brought about a drastic enhancement of the ferromagnetism in {\ZCT} while the grown films remained electrically insulating.
In particular, at a fixed Cr composition of $x = 0.05$, the ferromagnetic transition temperature {\Tc} increased up to $300\:{\rm K}$ at maximum due to the iodine doping from $\mbox{\Tc} = 30\:{\rm K}$ of the undoped counterpart, while the ferromagnetism disappeared due to the doping of nitrogen as a $p$-type dopant.
The observed systematic correlation of ferromagnetism with the doping of charge impurities of both $p$- and $n$-type, suggesting a key role of the position of Fermi level within the impurity $d$-state, is discussed on the basis of the double exchange interaction as a mechanism of ferromagnetism in this material.

\end{abstract}
\maketitle
%
%
Recently intensive efforts have been made for the search of novel ferromagnetic semiconductors{\cite{Dietl_Nmat03}}.
Up to the present, several materials have been reported to exhibit ferromagnetism above room temperature{\cite{Macdonald_Nmat05}}, but few of them have been confirmed to be ferromagnetic as an intrinsic nature of the material studied.
This is partly due to difficulties in excluding the contribution from possible precipitates of other compounds than the pure diluted phase.
In this regard, Cr-doped ZnTe is one of the promising candidates because the room-temperature ferromagnetism was observed in a crystal containing $20\:\%$ of Cr and its intrinsic nature was confirmed through magneto-optical measurements{\cite{Saito_PRL03}}.
However, the origin of ferromagnetism should be different from the carrier-induced mechanism as in Mn-doped III-V compounds{\cite{Macdonald_Nmat05}} because Cr ions are considered to be incorporated as Cr$^{2+}$ in II-VI compounds{\cite{Vallin_PRB70}} and to be electrically neutral.
Then a question arises how the additional carrier doping affects magnetic properties of this material.
Experimental studies addressing to this critical issue are expected to provide a substantial clue for understanding the mechanism responsible for ferromagnetism.
In our previous study{\cite{Ozaki_APL05}}, we reported the suppression of ferromagnetism in $p$-type {\ZCT} co-doped with nitrogen (N) as a dopant.
This result, which is against a na\"{\i}ve expectation from the analogy of (Ga,Mn)As, has brought our attention to the effect of doping of the opposite charge impurity.
Then, in the present study, we have investigated magnetic properties of {\ZCT} films co-doped with iodine (I), which is expected to act as an $n$-type dopant.
As a result, we found that the ferromagnetism was enhanced drastically due to the iodine doping in {\ZCT}.

%
An epitaxial layer of {\ZCT} was grown by conventional molecular beam epitaxy (MBE) method using solid sources of Zn, Cr and Te.
Iodine (I) was used as an $n$-type dopant and it was supplied as a compound of CdI$_2$.
A ZnTe buffer layer (thickness $\sim 700\:{\rm nm}$) was first grown on a GaAs (001) substrate to relax a large lattice mismatch and then a {\ZCT} layer (thickness $\sim 300\:{\rm nm}$) was successively grown on it.
The substrate temperature during the growth of the both layers was kept at $300^{\circ}{\rm C}$.
The growth of a ZnTe buffer layer and an undoped {\ZCT} layer was performed in the Te-rich condition while an iodine(I)-doped {\ZCT} layer was grown in the Zn-rich condition in order to suppress a possible formation of Zn vacancy due to the self-compensation effect.
The Cr composition $x$ in {\ZCT} layers was estimated using electron probe microanalyzer with a low acceleration voltage in order to probe only the {\ZCT} layer and the iodine concentration {\I} was measured using secondary ion mass spectroscopy (SIMS) with an ion-implanted ZnTe wafer as a reference.
In the present study, the Cr composition $x$ of {\ZCT} was fixed at around $x = 0.05$ and the iodine concentration {\I} was changed in the range of $10^{17} \sim 10^{19}\:{\rm cm}^{-3}$ by controlling the temperature of a CdI$_2$ $K$-cell.

%
The crystal structure of the grown films was examined by x-ray diffraction (XRD) and high-resolution transmission electron microscope (HRTEM) with a particular attention to check the presence of any secondary phase.
Figure 1 shows $\theta$--$2 \theta$ diffraction pattern of a typical {\ZCTfive}:I film, together with that of a ZnTe for reference.
There appeared only diffraction peaks of ZnTe $(00n)$ and GaAs $(00n)$ and any trace of other phases was not detected in a dynamic range larger than five order of magnitudes.
The diffraction peaks from the {\ZCTfive} layer did not split from those from the ZnTe buffer layer because there is little change in the lattice parameter due to the Cr incorporation{\cite{Ozaki_pssc04}.
Cross-sectional HRTEM image and selected-area electron diffraction (SAED) of {\ZCTfive}:I are shown in Fig. 2.
As shown in the figure, the lattice image exhibits a mostly single-crystalline zinc-blende structure, but there exist stacking faults along the $\left\{ 111 \right\}$ plane in some limited regions.
In SAED, intense spots in the hexagonal arrangement correspond to the zinc-blende structure, but additional weak spots were observed in one-third positions on the lines connecting the fundamental spots.
These extra spots are considered to reflect a triplet periodicity of the stacking sequence along the $\left\{ 111 \right\}$ plane in the stacking fault regions observed in the lattice image.
However, any apparent precipitates of other crystal structures were not detected both in the lattice image and the electron diffraction.
The result of TEM analyses of I-doped films is almost similar to that of undoped ones with the same Cr composition {\cite{Ozaki_pssc04,Ozaki_JSC05}} and the iodine doping causes little changes in the microscopic crystal structure of {\ZCT}.
\begin{figure}
\hspace*{-10mm} \vspace*{-10mm}
\includegraphics[keepaspectratio=true, width=9.5cm]{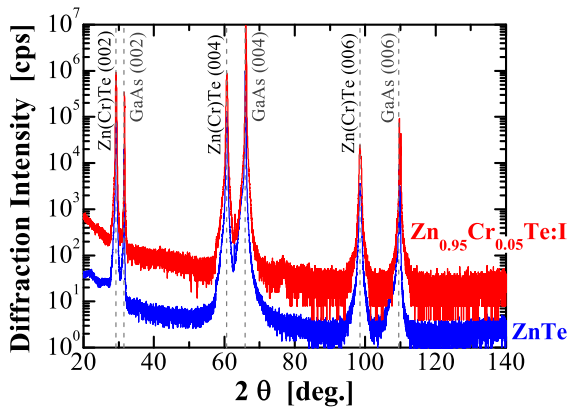}
\caption{\label{Fig1}
(color online) 
The $\theta$--$2 \theta$ XRD pattern of ZnTe and {\ZCTfive}:I films.
The iodine concentration of the I-doped film is $2 \times 10^{18}\:{\rm cm}^{-3}$ ($T_{\rm CdI_2} = 200\:^{\circ}{\rm C}$).
The vertical scale for the {\ZCTfive}:I film is shifted for clarity.
}
\end{figure}

\begin{figure}
\includegraphics[keepaspectratio=true, width=7.5cm]{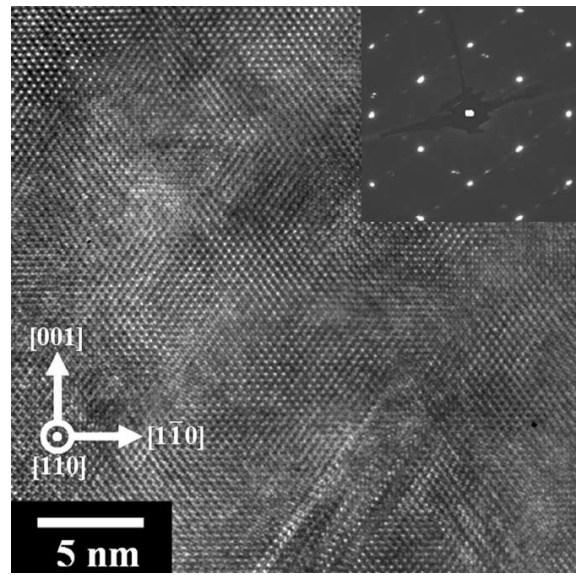}
\caption{\label{Fig2}
Cross-sectional HRTEM image of the same {\ZCTfive}:I film as shown in FIG. 1.
In the lower-right region in the image, stacking faults along the $\left\{ 111 \right\}$ plane were observed.
The inset shows the selected-area electron diffraction (SAED) pattern.
}
\end{figure}

%
The electrical measurement revealed that grown films of {\ZCT}:I remained insulating even with the highest doping concentration of the order of $10^{19}\:{\rm cm}^{-3}$.
This is in contrast to nitrogen(N)-doped {\ZCT}, which exhibited the $p$-type conductivity ranging from insulating to metallic behaviors depending on the balance between Cr and N concentrations{\cite{Ozaki_APL05}}.
It is known that $n$-type doping is exceptionally difficult in the host binary compound ZnTe, due to a kind of self-compensation effect{\cite{Chadi_PRL94,Marfaing_JCG96}}.
Indeed, there has so far been no achievement of $n$-type conduction in ZnTe using iodine as a dopant{\cite{Tanaka_JAP03}}.
An additional possible disadvantage in (Zn,Cr)Te is that electrons supplied from the donors would be trapped by the Cr impurity level within the gap.
Anyway, the $n$-type conduction could not be achieved in our {\ZCT}:I films, but it turned out that the iodine doping exerts a significant effect on magnetic properties.

%
The magnetization of grown films was measured using superconducting quantum interference device (SQUID) magnetometer with magnetic fields perpendicular to the film plane.
Figure 3 shows the magnetization versus magnetic field ($M$--$H$) curves of {\ZCTfive} films without and with the charge doping.
Compared to the hysteretic $M$-$H$ curve in the undoped film, the hysteresis loop was enlarged in the I-doped film ($\mbox{\I} = 2 \times 10^{18}\:{\rm cm}^{-3}$), while it disappeared in the N-doped film ($[{\rm N}] = 5 \times 10^{20}\:{\rm cm}^{-3}$).
At the same time, the saturation magnetization decreased in the N-doped film while it increased in the I-doped film.
The ferromagnetic transition temperature {\Tc} estimated from Arrott plot analysis was given by $\mbox{\Tc} = 300\:{\rm K}$ in the I-doped film and $\mbox{\Tc} = 30\:{\rm K}$ in the undoped film, while the N-doped film remained paramagnetic down to the lowest temperature of $2\:{\rm K}$.
%
\begin{figure}
\hspace*{-10mm} \vspace*{-10mm}
\includegraphics[keepaspectratio=true, width=9.5cm]{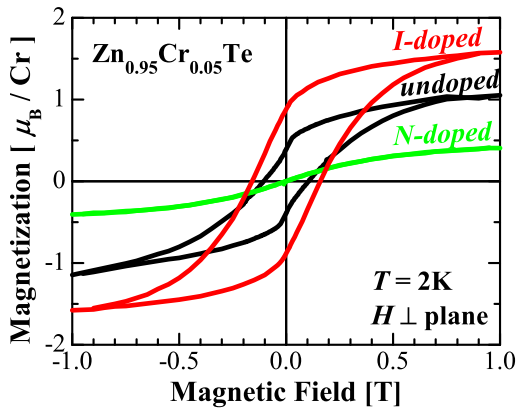}
\caption{\label{Fig3}
(color online) 
$M$--$H$ curve of undoped, N- and I-doped {\ZCTfive} films.
The measurement was performed with magnetic fields perpendicular to the film plane at $2\:{\rm K}$.
The doping concentration in the N- and I-doped films is $\mbox{\N} = 5 \times 10^{20}\:{\rm cm}^{-3}$ and $\mbox{\I} = 2 \times 10^{18}\:{\rm cm}^{-3}$ ($T_{\rm CdI_2} = 200\:^{\circ}{\rm C}$), respectively.
}
\end{figure}

The dependence of the ferromagnetic properties on the iodine concentration was investigated on a series of {\ZCT}:I films grown under a fixed Cr flux and different CdI$_2$ $K$-cell temperatures {\TCdItwo}.
Figure 4 shows the ferromagnetic transition temperature {\Tc} for this series of films, together with the paramagnetic Curie temperature {\thetaP} deduced from the Curie-Weiss plot.
In the figure, these critical temperatures are plotted as a function of the inverse of {\TCdItwo} and the scale of {\I} estimated from the SIMS measurement is given in the horizontal axis below.
With the increase of {\TCdItwo}, {\Tc} first increased from $30\:{\rm K}$ in an undoped film, and then reached the maximum of $300\:{\rm K}$ at $\mbox{\TCdItwo} = 200\:^{\circ}{\rm C}$ ($\mbox{\I} = 2 \times 10^{18}\;{\rm cm}^{-3}$) and then turned to decrease with the further increase of {\TCdItwo}.
On the other hand, {\thetaP} almost followed the increase of {\Tc} at low {\TCdItwo}, but it remained constant after reaching the maximum of $\mbox{\thetaP} = 320\:{\rm K}$ at $\mbox{\TCdItwo} = 200\:^{\circ}{\rm C}$.
It was found that {\Tc} was increased due to the iodine doping in other Cr compositions than $x = 0.05${\cite{Note_1}}.
(For example, at $x = 0.01$, {\Tc} reached $100\;{\rm K}$ with the doping of iodine at a concentration of $\mbox{\I} = 2 \times 10^{18}\;{\rm cm}^{-3}$, in contrast to $\mbox{\Tc} = 5\;{\rm K}$ in the undoped film.)
%
\begin{figure}
\hspace*{-10mm} 
\includegraphics[keepaspectratio=true, width=9.5cm]{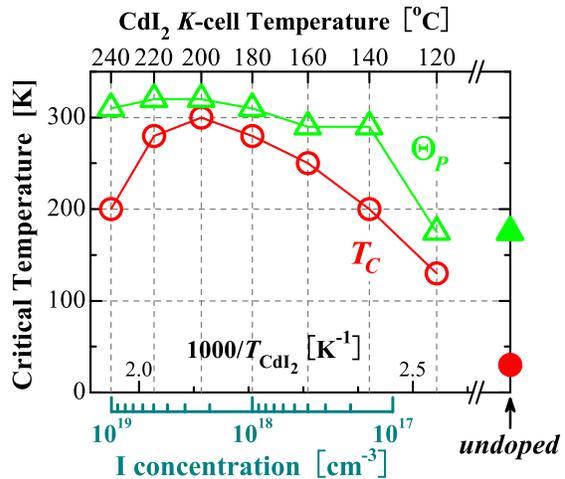}
\caption{\label{Fig4}
(color online) 
The ferromagnetic transition temperature {\Tc} and the paramagnetic Curie temperature {\thetaP} are plotted against CdI$_2$ $K$-cell temperature {\TCdItwo}.
The Cr composition $x$ is almost constant at around $x = 0.05$ and the scale of the iodine concentration {\I} is given in the horizontal axis below, based on an estimate by SIMS measurements.
{\Tc} and {\thetaP} of an undoped {\ZCTfive} are plotted on the right edge of the figure by closed symbols.
}
\end{figure}

%
Magnetic circular dichroism (MCD) measurements were performed in order to dinstinguish the magnetic properties which are intrinsic in the pure diluted phase of (Zn,Cr)Te from those of possible precipitates of other compounds.
Figure 5(a) shows the MCD spectra of ZnTe and {\ZCTseven}:I at $2\:{\rm K}$, $1\:{\rm T}$.
In ZnTe, a sharp peak appeared at $2.38\:{\rm eV}$, corresponding to the intrinsic Zeeman splitting at the $\Gamma$ point.
In {\ZCTseven}:I, on the other hand, a broad negative band appeared in the lower-energy side of a similar sharp peak at $2.38\:{\rm eV}$.
These characteristic MCD spectra were also observed in undoped {\ZCT}{\cite{Saito_PRL03,Kuroda_STAM05}}.
Though the origin of this broadening has not been yet specified{\cite{Note_2}}, this broad MCD band is considered to be related to the enhanced magneto-optical effect at the $\Gamma$ point{\cite{Saito_PRL03}}.
In Fig. 5(b), the MCD intensities at a fixed photon energy are plotted against the magnetic field and compared with the magnetization measured by SQUID.
As shown in the figure, the magnetic-field dependence of MCD coincides well with the $M$--$H$ curves at the respective temperatures, although there is a slight difference in the ratio between the MCD intensities and the magnetizations at different temperatures.
This coincidence corroborates that the Cr $d$ electrons exhibit the exchange interaction with the $sp$ electrons at a critical point of the energy band;
that is, the observed ferromagnetic behaviors are intrinsic properties of (Zn,Cr)Te with Cr incorporated substitutionally in the Zn site, not coming from precipitates of other Cr compounds.

\begin{figure}
\hspace*{-10mm} \vspace*{-10mm}
\includegraphics[keepaspectratio=true, width=9cm]{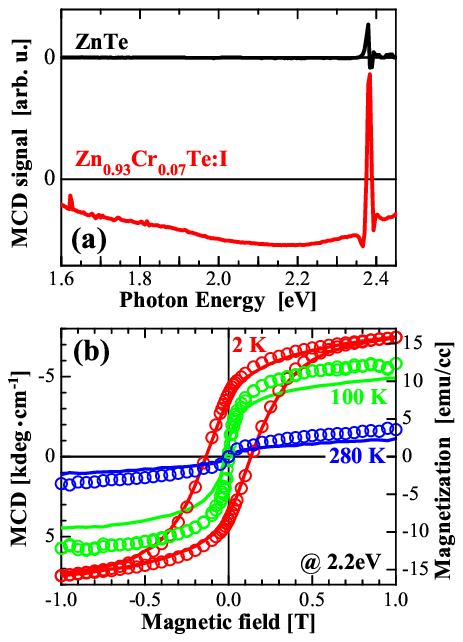}
\caption{\label{Fig5}
(color online) 
(a) The MCD spectra of ZnTe (without I doping) and {\ZCTseven}:I ($\mbox{\TCdItwo} = 200\:^{\circ}{\rm C}$, $\mbox{\Tc} = 280\:{\rm K}$) films at $2\:{\rm K}$, $1\:{\rm T}$.
The measurement was performed in the transmission mode under magnetic fields perpendicular to the film plane (Faraday configuration).
The vertical scale is the same for the two spectra but the origin is shifted for clarity.
(b) The field dependence of the MCD signal intensities at a photon energy of $2.2\:{\rm eV}$ (represented by lines), and the magnetization measured by SQUID (represented by marks).
}
\end{figure}

%
The direction of change due to the doping is opposite to the case of (III,Mn)As, in which the ferromagnetism is mediated by holes and therefore {\Tc} is scaled by the hole concentration{\cite{Macdonald_Nmat05,Edmonds_APL02}}.
Since the origin of ferromagnetism in (Zn,Cr)Te is considered to be different from the carrier-induced mechanism in (III,Mn)As, another explanation is needed for the doping effect.
Concerning the origin of ferromagnetism in Cr-doped II-VI compounds, both the superexchange interaction {\cite{Blinowski_PRB96}} and the double-exchange interaction {\cite{Sato_SSC02}} have been proposed.
According to the first-principle calculations by Sato {\it et al.}{\cite{Sato_SSC02}}, the Cr $3d$ electrons form the non-bonding $e$ state and the anti-bonding $t_a$ state within the band-gap of ZnTe and the Fermi level {\EF} is located in the middle of the upper $t_a$ state.
Electron hopping between this partially-occupied $t_a$ state stabilizes the ferromagnetic alignment of Cr spins (double-exchange mechanism{\cite{Akai_PRL98}}).
Since the strength of this ferromagnetic double-exchange interaction is dependent on the occupation of the $d$-state, the shift of {\EF} due to the charge doping is expected to affect the ferromagnetic properties.
The suppression of ferromagnetism in $p$-doped (Zn,Cr)Te is explained well by this picture{\cite{Ozaki_APL05}}; that is, the downward shift of {\EF} due to the doping of an acceptor impurity reduces the electron density in the Cr $3d$ state, resulting in the suppression of the double-exchange interaction.

On the other hand, {\EF} is expected to shift upward upon the doping of a donor impurity.
In order to explain the observed enhancement of ferromagnetism in the I-doped films, it should be assumed that {\EF} in the nominally undoped crystal is deviated below from the peak of the density of states of the $t_a$ state, presumably due to a native Zn-vacancy or other defects induced by the Cr incorporation.
Assuming this initial position of {\EF}, the upward shift of {\EF} due to the iodine doping increases the electron density at {\EF}, resulting in the enhancement of the ferromagnetic interaction.
This picture, based on the double-exchange mechanism, could give a qualitative explanation for the effect of the charge doping, at least for the direction of change observed in the experiments{\cite{Note_3}}.
Further investigation will be needed to examine this picture in detail, including a quantitative analysis and an experimental probe of the change in the valence of Cr ions{\cite{Vallin_PRB70}}.
Anyway, the observed systematic and spectacular change of ferromagnetism demonstrates that the position of the Fermi level within the localized $d$-state of the magnetic impurity plays a key role and the ferromagnetic properties can be manipulated by the doping of charge impurities through the shift of the Fermi level{\cite{Reed_APL05}}.

%
In summary, we reported a significant enhancement of ferromagnetism in {\ZCT} due to the doping of iodine as an $n$-type dopant.
At a fixed Cr composition $x = 0.05$, {\Tc} increased up to $300\:{\rm K}$ at maximum from $T_{\rm C} = 30\:{\rm K}$ in the undoped counterpart, while the ferromagnetism was not observed in $p$-type samples doped with nitrogen.
The systematic correlation of ferromagnetism with the doping of charge impurities of both $p$- and $n$-type, with the variation of {\Tc} extending over a wide range of $0 \sim 300\:{\rm K}$, demonstrates a key role of the position of the Fermi level within the impurity $d$-state.
These experimental findings are expected to exploit the possibility of manipulating ferromagnetism by external charge control.

%
The authors would like to thank Prof. K. Kadowaki for support in SQUID measurement.
Thanks are also due to Prof. Y. Masumoto for support in TEM analysis and due to Dr. T. Ohshima (Japan Atomic Energy Agency) for the preparation of the SIMS reference samples.
This work has been partially supported by the Grant-in-Aid for Scientific Research in Priority Areas `Semiconductor Nanospintronics' and 21st Century Center of Excellence (COE) Program `Promotion of Creative Interdisciplinary Materials Science for Novel Function' from the Ministry of Education, Culture, Sports, Science and Technology, Japan (Monbukagakush\^o). \\
%
%
%
%

\begin{thebibliography}{99}
\bibitem{Note_1}
The additional experiments of the post-growth annealing revealed that the ferromagnetic properties including {\Tc} did not change upon annealing at slightly higher temperatures ($320 \sim 340\:^{\circ}{\rm C}$) than tne growth temperature.
\bibitem{Note_2}
From the similarity in the spectral shape with the optical absorption and photo-excited EPR{\cite{Godlewski_JPC80}}, this MCD band might possibly be assigned to the transition from the valence band to the Cr impurity level{\cite{Marcet06}}.
\bibitem{Note_3}
The decrease of {\Tc} after reaching the maximum of $\mbox{\Tc} = 300\;{\rm K}$ might also be explained as a result of the shift of the Fermi level within the Cr $3d$ band; with the increase of {\I}, the Fermi level exceeds the point of the maximum density of state in the $t_a$ level, which results in the decline of the double-exchange interaction.
Discrepancies between {\thetaP} and {\Tc} in this range, which were also observed in undoped {\ZCT} with high Cr compositions{\cite{Ozaki_JSC05}}, suggest the inhomogeneity of the magnetic properties.
\end{thebibliography}
%

%
\end{document}